# MACHINE LEARNING ALGORITHM FOR NLOS MILLIMETER WAVE IN 5G V2X COMMUNICATION

Deepika Mohan[1], G. G. Md. Nawaz Ali[2] and Peter Han Joo Chong[1]

[1]Department of Electrical and Electronics Engineering,
Auckland University of Technology, Auckland 1010, New Zealand
[2]Department of Applied Computer Science,
University of Charleston, WV 25304, USA

## ABSTRACT

*The 5G vehicle-to-everything (V2X) communication for autonomous and semi-autonomous driving utilizes the wireless technology for communication and the Millimeter Wave bands are widely implemented in this kind of vehicular network application. The main purpose of this paper is to broadcast the messages from the mmWave Base Station to vehicles at LOS (Line-of-sight) and NLOS (Non-LOS). Relay using Machine Learning (RML) algorithm is formulated to train the mmBS for identifying the blockages within its coverage area and broadcast the messages to the vehicles at NLOS using a LOS nodes as a relay. The transmission of information is faster with higher throughput and it covers a wider bandwidth which is reused, therefore when performing machine learning within the coverage area of mmBS most of the vehicles in NLOS can be benefited. A unique method of relay mechanism combined with machine learning is proposed to communicate with mobile nodes at NLOS.*

## 1. INTRODUCTION

Significant advancement has been made in the field of vehicular communication in recent years. Information about live traffic updates, warnings, safety messages and even more information are exchanged through the communication node among vehicles and Road side units (RSU) [12]. All these communications between vehicles and network are done as the vehicles communicates through cell phone towers for information about traffic and routes. Every mobile node communicates with each other to avoid accidents and to calculate the speed of their neighbours, hence from this it is clear that the vehicular communication and mobility are approaching for a new era. The 5G provides a higher speed with lower latency which creates a revolution in the digital world of wireless technology and the advancements in 5G has benefited in the improvements in vehicular networks for autonomous driving [11]. Ultra-high reliability is provided for many applications that utilize 5G as it involves new technology with a usage of high frequency and antenna improvement. The vehicle-to-everything (V2X) plays an important role in improving high vehicle utilization, supporting accident free transportation thus providing zero emission vehicles which are more efficient [16]. The major use of 5G technology in future is V2X communication where the vehicles communicate with its surroundings to get information from outside world. For basic security and non-safety applications the V2X technology is operated below 6 GHz [26]. Many V2X applications make use of the Millimeter Wave (mmWave) band as it functions below 6GHz with higher data rate, increased network capacity and bandwidth. The major drawback mmWave faces is its coverage area as it suffers a higher penetration loss hence it cannot transmit signal when obstructed by buildings, therefore mmWave cannot send information to vehicles in NLOS. The mmWave coverage area can be increased either by deploying mmWave Base Station (mmBS) or designing an antenna in such a way which will transmit information to vehicles using radio frequency [13]. But a suitable solution for overcoming this drawback is by using Machine Learning (ML); through which mmBS understands and learns about its environment and transmits data to vehicles in NLOS range [6]. Even though mmWave have more advantages in the communication field, it suffers a big loss due to shadowing, blockages, fast fading and path loss [5]. More number of approaches is proposed for these drawbacks but for predicting blockages and transmitting messages is still a challenge in mmWave, which will be focused in this paper. Therefore in a particular coverage area if the broadcast information such as basic safety messages (BSM) are transmitted to all vehicles then only the vehicles at LOS receives the information but other vehicles in NLOS cannot receive the message from mmBS. Many researches are going on to overcome this problem in mmWave. One solution is by making use of ML so that all the vehicles can communicate with the mmBS without any interruptions as ML trains the base station to identify blockages and transmits data to NLOS nodes in terms of automatic and semi-autonomous driving which is the main motive of this paper. From the best of our knowledge none of the existing works proposed this idea and shown any result. Hence the organizational structure of this paper is as follows, this paper introduced the functions and applications of mmWave and Machine Learning in section 1. Section 2 discusses the related works which helps the reader to understand the different methodologies used in mmWave and V2X Communication using ML. The System Architecture and its work functions are summarized in Section 3 and Section 4 discusses the RML algorithm and the experimental results. Finally Section 5 concludes the overall work along with the limitations of RML and its future developments.

## 2. RELATED WORKS

For improving the efficiency of road safety, traffic and infotainment options in vehicles, the V2X communication has been identified as the most suitable technology between road infrastructure and vehicles [25]. The V2X provides better QoS and coverage when compared to the other short-range dedicated communication. The mobile nodes handle more amount of data due to crowding and video streaming on traffic and entertainment applications with frequent internet usage [7]. Hence for this each automatic application requires good reliability with reduced latency which is embedded within individual packet preference level. In order to lower the pressure of complexity in vehicular applications, the ML is developed for V2V communication in which every transmitter acts as an agent whose decision is based on the observation from the surroundings. Even though ML and 5G are two different fields but on combining both together better results are obtained for solving various application problems. The complexity of design and procedures in V2X can be handled by data learning, modifying and replacing the rule list with ML routines which learns readily from previous stored data. The problems in coverage probability on spectrum location are differentiated into small parts by using Poisson Point Process and Cox Point process [10]. For the vehicular Fog computing methodology by using Q-learning provides higher reward as the system reaches a stable state faster when compared to other existing methods [27]. Considering the expected high traffic demand of vehicular networks, the spectrum resources of mmWave are used more efficiently for spectrum sharing mechanism. Therefore a model based on sensing is developed on mmWave spectrum for vehicular networks [24]. For constructing this model innovatively, an algorithm based on beam alignment was proposed using the temporal correlation. During the implementation of this method, the sensing outputs generated in different sensing time are affected by various environmental conditions.

On utilizing the capable solutions like the beam recovery, tracking and alignment, the higher data rate of 10 to 100 mbps and lower latency of 10ms to 100 ms in mmWave can be used by vehicular communication [19]. In order to overcome the limitations of hardware in mmWave, the beam forming mechanism is proposed in which the antenna weight is restricted to phase shifts of lower resolution and the best transreceiver pairs are selected [21]. The training process helps to detect the capable beam pairs that increase the standard of the link using numerical algorithm [22]. The ML designs can utilize the uplink signal gathered at the base station to learn about the mapping structure related to the outdoor scenario [9]. Even though there are many solutions supporting mmWave communication but the actual problem lies in the system design [8]. It suffers a higher penetration loss when blocked by obstacles and hence results in inaccurate beam selection. There are some methods that are time consuming and unscalable for 5G implementation but these approaches cannot detect regular pattern of traffic and blockages. An approach of contexted multi- armed bandit was developed that works on the network information based on the vehicles arrival direction which uses the mmBS to automatically learn and understand from its environment. Many NLOS scenarios have dissimilar amplitude giving rise to functions consisting of various local optima and hence by improving channels of NLOS nodes certain changes and extensions are proposed in Nelder-Mead beam based training technique [14]. Therefore the starting point, of the training mechanism is derived as in Equation (1) as,

$$P_{1,nc}^2 = (p_{1,nc}^2, q_{1,nc}^2), n_c \in [1, N_c] \quad (1)$$

$$p_{1,nc}^2 = 2 \text{ x } (2 \text{ x } P_{opt,nc}^1 - 1) - 1 \quad (2)$$

$$q_{1,nc}^2 = 2 \text{ x } (2 \text{ x } q_{opt,nc}^1 - 1) - 1 \quad (3)$$

where $N_c$, is the output optima which is defined as, $P_{opt,nc}^2 = (p_{opt,nc}^2, q_{opt,nc}^2)$, declared in descending order which is derived from Equation (2) and (3). The ML algorithm based on Manhattan Poisson line process for street model is defined as a better solution for differentiating blockages in mmWave urban networks [17]. In order to analyze the blocking assistance the following equations are calculated for simulating triangles as in Equation (4).

$$\frac{B_C^m - \Omega_m - \frac{w_v^2}{2}}{B_C^m} = \frac{H_{BS} - H_L}{H_{BS} - H_S} \quad (4)$$

Therefore from Equation (5), the final value of blocking assistance is derived as,

$$B_C^m = \frac{\Omega_m + \frac{W_v^l}{2}}{[\frac{H_{BS} - H_L}{H_{BS} - H_S}]} \quad (5)$$

For a group of vehicles travelling on the same lane, some mobile nodes experience fading if the length of the group is long and hence a relay option is best suited to overcome this path loss [3]. The most efficient method for penetration loss problems in mmWave is to utilize ML. A relay mechanism based on soft-information was delivered for vehicular networks which use the estimate and forward strategy in which the minimum mean square error of the signal received is obtained with less complexity which achieves a good trade off among vehicular networks [1]. The DSRC and LTE-V2V nodes use the first relay system which provides a performance gain of 91% with respect to the communication distances [2]. For signal blocking in a direct link a deep learning model was formulated to solve the challenges in optimal relay selection [4]. This method of prediction is utilized in the proposed research in terms of relay selection but using RL.

## 3. SYSTEM MODEL

Since more number of solution were formulated for sending information to NLOS vehicles, these solutions may sometimes make the system complex due to increased system specification in mmWave links. In this paper, we propose Relay using Machine Learning (RML) algorithm to train the mmBS for identifying the blockages and to learn about the best direction and location for broadcasting the messages to vehicles at NLOS. A relay mechanism is used for transmitting information to vehicles that cannot communicate with the mmBS. This proposed ML algorithm chooses a relay that is capable of transmitting data to all the vehicles at NLOS with strong signal strength and minimum congestion. This kind of Relay mechanism utilizing ML is new in mmWave 5G V2X communication as the BS learns about its environment through observations and transmit information to vehicles at NLOS through a single hop relay depending on the position of blockages. Instead of deploying more number of BS to broadcast messages, it is easier for formulating ML on mmBS and make it learn about its surroundings in order to transmit data accordingly which saves much time, cost and energy. An urban scenario is considered for the research where there are rapid changes in traffic and the network is often interrupted by permanent and temporary blockages. Fig. 1 shows the design scenario of the proposed research. The mmBS which is deployed has a coverage range of 300 meters and the vehicles on different lanes are subjected to blockage effect. The vehicles closer to mmBS are free from blockages as they are at direct LOS with the BS. But the vehicles at a distance are obstructed by blockages is the NLOS node that does not receive any data from the mmBS.

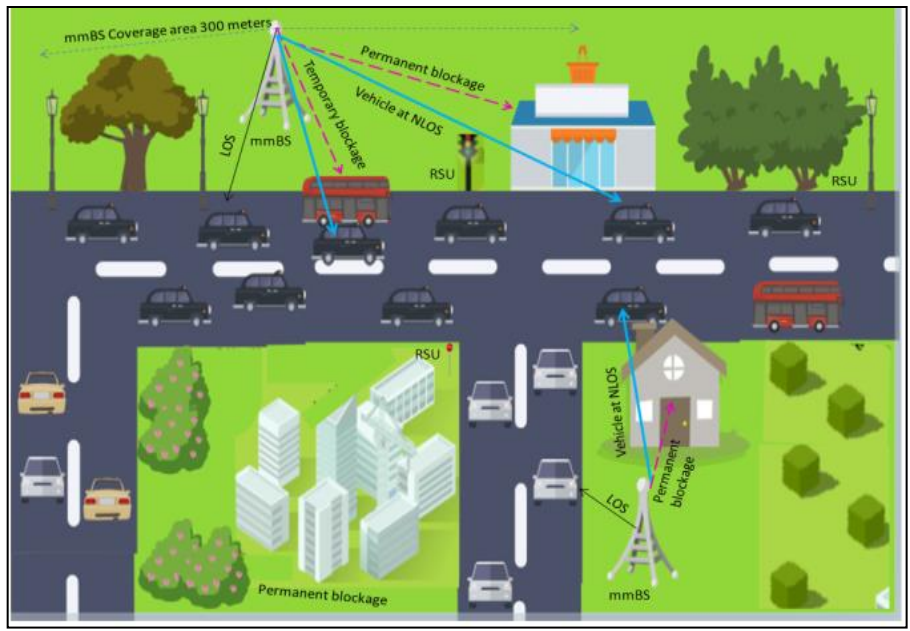

Fig. 1. Architectural design scenario.

### 3.1. Identification of Blockages

As the mmWave is sensitive to blockages, it leads to severe shadowing and produces various characteristic of path loss between LOS and NLOS links. In order to reduce this effect, the blockage is identified [18]. Using the RML the mmWave identifies the blockages. Initially the mmBS broadcasts signal through antenna and the blockages are identified if the radiation pattern is reflected back from its path. After a deep research analysis on the temporary blockage an assumption is made in this paper that the maximum height of vehicle is 16 feet, and only large vehicles like bus, truck and semi-trucks are considered as temporary blockages. Hence a threshold value of 16 feet is taken to differentiate blockages as permanent or temporary. Once when the radiation pattern is reflected below the threshold the mmBS identifies it as temporary blockage and if the reflection occurs greater than the threshold it determines it as permanent blockage. In the case of temporary blockages the obstacle itself acts as a relay to transmit information to NLOS nodes, but for permanent blockages the BS calculates the distance between itself and the barrier along with the direction and hence stores the detail for future use. This stored data on the position of blockages is updated at regular intervals of time which makes the mmBS understand the best location and position to communicate with vehicles at NLOS. Fig. 2 shows the distance estimation for identifying the blockage location.

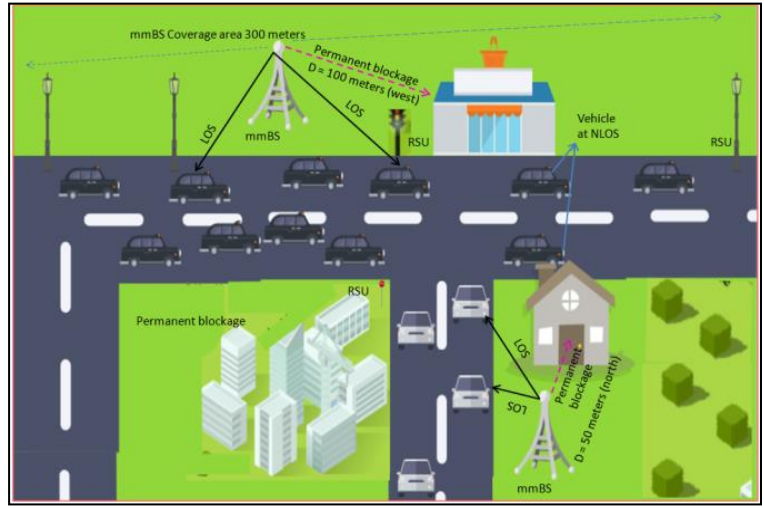

Fig. 2. Identification of the location of blockage based on distance estimation.

## 3.2. Efficient Relay Selection

The mmBS identifies the location of LOS and NLOS vehicles in its coverage area through V2V communication. Once when a LOS vehicle communicate with the BS, all the data about the LOS is gathered by the mmBS, hence the BS confirms the position of NLOS node and chooses an efficient relay for transmission of information in the following manner [20]. In handoff, a BS targets a particular mobile node and manages data delivery through relay technique. From Equation (6), the handoff algorithm is formulated at time t, with N, number of flow at m (constant) and hence the estimation state is given as,

$$N_n = N + N_r - N_t - m \qquad (6)$$

$$N_r = t_s * \frac{number of arrival}{T_e} \qquad (7)$$

$$N_t = t_s * \frac{number of departure}{T_e} \qquad (8)$$

where $t_s$ the next sampling time and $T_e$ is the estimated time as in Equation (7) and (8). Using RML the mmBS chooses the suitable relay based on the following conditions,

a. Since most vehicles uses GPS receiver system, each node in the network gets its position in real time.

b. The link metric is calculated by exchanging information about the vehicles position through control messages which is stated in Equation (9) as,

$$\text{Link metric} = \ln[1 - \prod_{V_{Ln} \in Path}(1 - P_{VLn})] \quad (9)$$

c. Based on the information exchanged the Relay calculation is performed with the metrics choosing Relay path as calculated in Equation (10).

$$\text{Path} = \{V_{L1}, V_{L2}, \ldots \ldots V_{Ln}\} \quad (10)$$

Here $V_{L1}, V_{L2}, \ldots \ldots V_{Ln}$ are different suitable LOS relay nodes. Hence the RML algorithm aims to choose a relay based on distance and position which is defined in Equation (11), (12) and (13).

$$P_{path} = 1 - \prod_{V_{Ln} \in Path}(1 - P_{VLn}) \quad (11)$$

$$P_{path} = \ln[1 - \prod_{V_{Ln} \in Path}(1 - P_{VLn})] \quad (12)$$

$$\sum_{V_{Ln} \in Path} \ln P_{VLn} = \sum_{V_{Ln} \in Path} LinkMetric_{VLn} \quad (13)$$

## 4. RML ALGORITHM

Machine learning algorithm is used in the beam to direct the messages and performs relay mechanism so that all the vehicles make use of the mmWave technology. On performing Beam selection or switching operations there is fewer guarantee that all the data transmitted reaches the mobile node without any delay but by using relay mechanism the data reaches the nodes at a faster rate without any packet loss. In this scenario all the network operations and functions are built on the same data foundations which will make the system simpler [15]. The advantage of RML is that the solution to an issue can be learnt directly from the data generated as this algorithm runs on the mmBS which broadcasts information to the nodes in its coverage area. The background of RML algorithm is the Reinforcement Learning which is formulated based on the correlation among the location of NLOS vehicles and the action taken by the mmBS based on Relay mechanism is the key for future decision.

### 4.1. Detailed Description

In detail, our proposed RML algorithm work function is as follows (see Algorithm 1): First during the initialization process the trained model accepts the input broadcast messages $B_{mn}$, height, distance and $V_h$. The mmBs coverage area is assigned to 300 meters and hence the system observes the environment based on regular traffic patterns. The model checks for permanent and temporary blockages based on the estimated threshold $\varepsilon$.

**Algorithm 1** Pseudocode of RML algorithm.

1: Input: $B_{mn}$, h, D, Ref, angle (θ) and $V_h$
2: Initialize vehicles: $V_h = Vh_1, Vh_2, \ldots \ldots \ldots Vh_n$
3: Initialize LOS nodes: $V_L = V_{L1}, V_{L2}, \ldots \ldots \ldots \ldots V_{Ln}$
4: Initialize NLOS nodes: $V_{NL} = V_{NL1}, V_{NL2}, \ldots \ldots \ldots \ldots V_{NLn}$
5: Output: Ar (antenna radiation), $R_t$
6: Initialize Model = model (input, output)
7: State = Observation

8: **if** State random.rand () $<\varepsilon$
9: Select action temporary blockage
10: **else** Take action $>\varepsilon$
11: Action = D + Ref + θ //Direction of permanent blockage
12: Return action
13: **end if**
14: Exchange $D_v, V_L, V_{NL}$ // vehicles exchange information about its distance and information about its neighbour nodes.
15: Calculate $V_D = D_r - D_v$ // distance between the vehicle and the blockage is calculated
16: State S = $V_D$ + + // update the state
17: Select Relay node $D_v \leq D_r$
18: **if** $(D_v(V_{L1}) \leq D_v(V_{NL1}))$ // calculates the distance between vehicle1 (LOS) with vehicle1 (NLOS)
19: Transmit Bm1 = $V_{NL1}$ // Send received broadcast message from vehicle1 (LOS) to vehicle1 (NLOS)
20: **else** $(D_v(V_{L2}) \leq D_v(V_{NL1}))$ // calculates the distance between vehicle2 (LOS) with vehicle1 (NLOS)
21: **end if**
22: Update $D_r = D_v$ // update the best distance between vehicles at LOS to NLOS based on the direction of permanent blockage.
23: S = np.array (a[0]) // Select States
24: S+1 (new state) = np.array (a[n]) // Select new States for replay
25: Q = self.model predict (states)
26: Q-new = self.model predict (new_states)
27: Replay_distance = len (replay)
28: Target = Q[i]
29: Target [action_r] = reward_r
30: **while True,**
31: Total reward = reward
32: S = S+1
33. State_s, action_a, reward_r, done_r = =replay [i] // To construct training set
34: **for** each $D_r$ // identify the best suitable distance of data transmission
35: Repeat Step 2
36: **end for**
37: $B_m$= $V_L$ = $V_{NL}$ // All the vehicles at both LOS and NLOS receives broadcast information from the mmBS
38: **end**
39: **Return**

When $\varepsilon$ is higher than 16 then the action is taken with respect to distance D, Ref and θ to estimate the location of the blockage. In the mean time, the vehicles $V_L$(LOS) and $V_{NL}$(NLOS) exchange information about each other and the overall data is termed as $V_h$. The mmBS calculates the distance between $D_r$ and $D_v$ to predict $V_D$ and hence the state S is ready for update. The Relay node is selected based on the distance between $D_v$ and $D_r$, hence initially the distance between the LOS node $V_{L1}$ and NLOS node $V_{NL1}$ is calculated and if the distance are near then $B_{m1}$ is transmitted to $V_{NL1}$ else the next LOS node $V_{L2}$ is compared until a suitable relay is selected. In order to transmit data to $V_{NL1}$ the target Q[i] is chosen which is the relay node and the action_r is performed and for every action a reward_r is obtained. Finally $B_{mn}$ is transmitted to $V_{NL}$ through $V_L$ with a new state S+1. The time complexity of RML algorithm measures the speed at which this algorithm performs its operations for the given input size n. This proposed RML algorithm has two types of time complexity: training and run time complexity. When the amount of data is larger, the time complexity for training the mmBS is represented as in Equation (14).

$$\text{Training time complexity} = O(n*\log(n)*d*k) \quad (14)$$

where n is the number of training examples which is chosen from the environment using RL and d is the directional dimensionality of blockages, and k is the number of available LOS relay nodes. The complexity in run time is expressed with respect to the traffic pattern and the neighbour nodes which is represented in Equation 15 as,

$$\text{Run time complexity} = O(\text{depth of traffic}*k) \quad (15)$$

where the depth of traffic denotes the overall nodes in the coverage area. In order to decrease the complexity of vehicles the action space is considered when a maximum update is reached the model resets itself without erasing the data from its memory.

### 4.2. Simulation Setup

The scenario is simulated for the terrain area with dimension 300m X 300m where the transmission range of mobile nodes and eNB are 100m. Table 1 briefs the important parameters used for simulation.

Table 1. Simulation parameters.

| Simulation Parameter | Values |
|---|---|
| Simulator tool | ns-3 |
| Machine Learning | Reinforcement Learning |
| Terrain Dimension | 300m x 300m |
| No. of eNB's | 1 |
| Position of eNB (x, y) | (55,55) (115,115) (175,175) (235,235) (295,295) for blockages (2, 4, 6, 8, 10) |
| Mobility model of eNB | Constant position model |
| No. of vehicles | 10, 20, 30, 40, 50 |
| Mobility model for vehicles | Random waypoint model |
| Speed of vehicles | 15 Min = 0.1 Max = 15 |
| No. of Relay | 1 |
| No. of Blockages | 2, 4, 6, 8, 10 |
| Propagation Model | 3GPP Propagation model |
| Building width | X = 50m, Y = 50m |
| Building height | Z = 10m (32 feet) |
| Channel Model | 3GPP Channel model |
| Transmission radius | 100m |
| Bandwidth of mmWave signal | 200 MHz |
| Transmit power | 30 dBm |
| Packet size of messages | 1024 bytes |
| Data Rate | 100 Gb/s |
| MTU | 1500 bytes |
| Interpacket arrival | 200ms |
| Queue size | 100 |
| Simulation Time | 50Seconds |

The number of blockages is varied within the coverage area based on the location of eNB which is deployed using constant position model. The mobile nodes are placed in the simulation setup using the Random waypoint model in which the speed of vehicle is set as 15 (constant) with a pause time of 5 seconds after which the vehicle takes new direction. An assumption is made in this simulation scenario that when a vehicle hits the blockage or the borders of the coverage area it bounces back and takes a new position as the collision is ignored in this setup because the main concentration of our work is relay selection and transmission of information to vehicles at NLOS. Here the maximum number of obstacles are given as 10 which is designed using the 3 GPP building model with a height of the obstacle, $z = 10$ meters. The simulation time is set as 50 seconds but the actual simulation runs for 45 minutes which is the time taken for training and run time purposes. Fig. 3 shows the design of the simulated network. For the location of permanent blockages a 3GPP propagation model is used to calculate the co-ordinates of x and y.

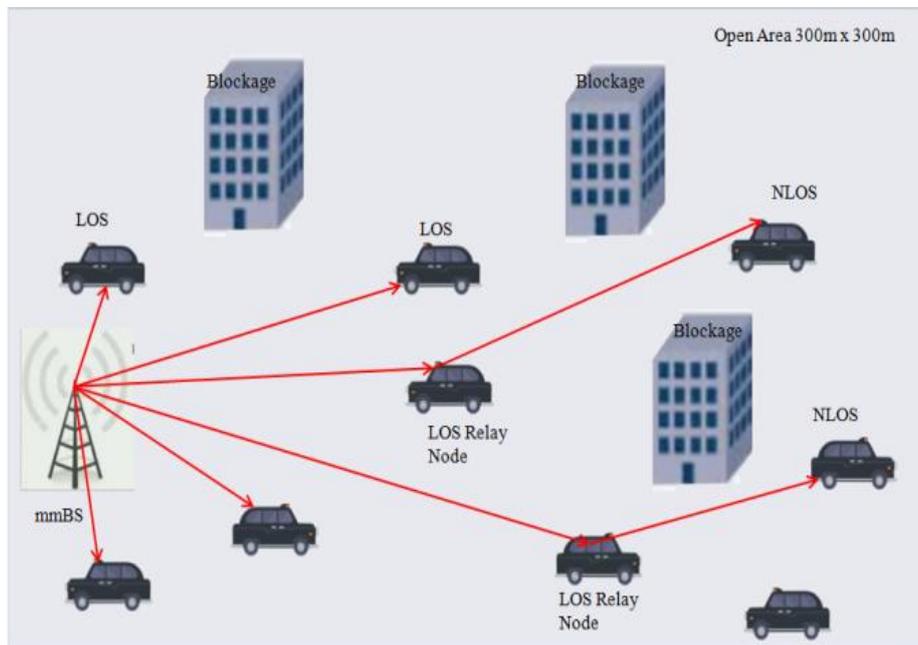

Fig. 3. Layout of the Simulated Network.

### 4.3. Experimental Results and Discussions

From the formulated RML algorithm the simulation is done using the specifications of the parameters. The results are obtained on comparing the simulation scenario using RML and without using RML. For the model without RML, the simulation is done with normal mmWave design which is prone to blockages and penetration loss. Hence the simulation is performed under 3 stages and the results are obtained in terms of throughput, latency and PDR. First the mobile nodes are kept constant at 20 and the blockages are varied from 2, 4, 6, 8 and 10 for the simulation time of 50 seconds. Fig. 4 shows that the average latency is less than 0.05 ms for the simulation that uses RML with 10 obstacles but the latency increases to 0.7 ms for the model without using RML.

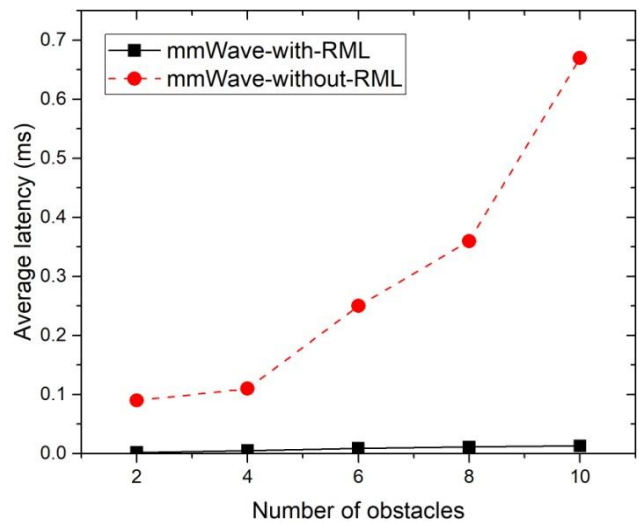

Fig. 4. Average latency comparision under different number of blockages for mmWave with and without RML.

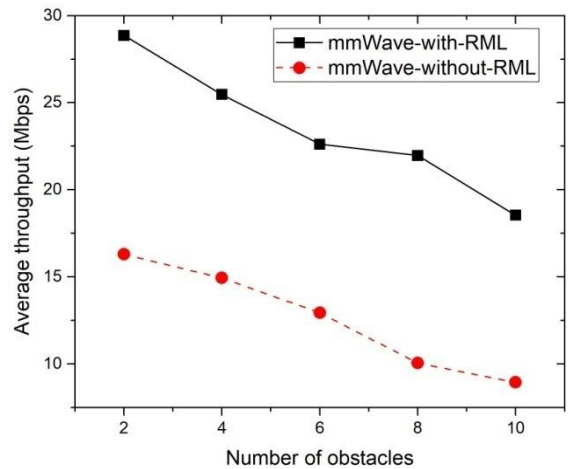

Fig. 5. Average throughput evaluation for mmWave with and without RML.

The throughput is calculated by varying the blockages and from Fig. 5 the throughput is 27.44 mbps when using RML but it is only 18 mbps when RML is not used. Hence the performance of the model without RML is low. Fig. 6 shows the result of PDR for constant mobile node where the performance is good as it reached 100% when using RML in mmWave than without using RML.

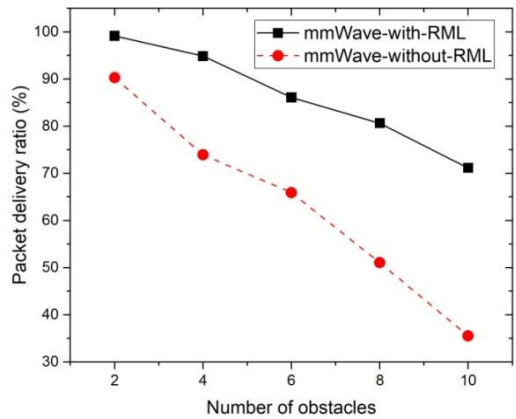

Fig. 6. PDR performance of mmWave through RML and without RML for constant mobile nodes.

In the next stage of the simulation the blockages are made constant at 10 and the vehicles are varied and from the result obtained in Fig. 7, it is clear that for maximum blockages the latency output using RML is at 0.1ms but for model without using RML it goes beyond 1.6ms.

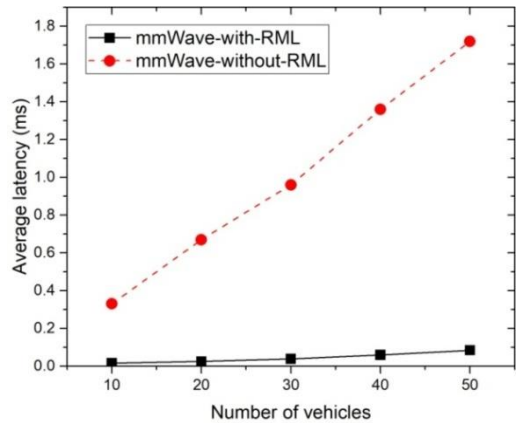

Fig. 7. Average Latency evaluation for Blockage = 10 by varying vehicles in mmWave with and without RML.

The Fig. 8 shows throughput result in which the model using RML outperforms the other.

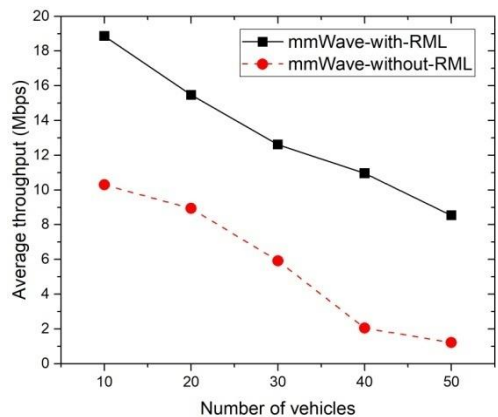

Fig. 8. Average throughput comparision graph for number of vehicles at blockage = 10 in mmWave with and without RML.

For the number of blockages as 10 and vehicles as 10 the PDR is 75% for the scenario using RML where as it is only 50% for the output without using RML as in Fig. 9. From the obtained result it is clear that on using RML as the blockages increases the throughput and the PDR also rises with a fall in latency, therefore the proposed RML confirms that this kind of solution can have the ability to solve real-world applications.

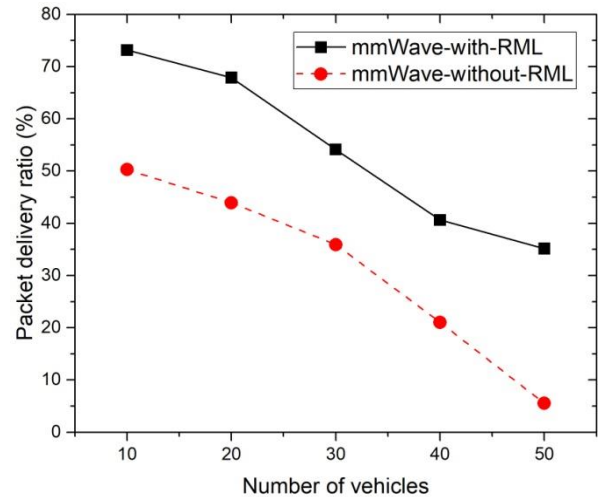

Fig. 9. PDR perfromance for mmWave with and without RML.

## 5. CONCLUSION

In this paper, we address the drawbacks in mmWave due to blockages from buildings and vehicles in real-time urban scenario. To this aim, we propose RML, a single hop relay mechanism based on Reinforcement learning which trains the mmBS to identify the location of blockages and to select a relay node based on the distance calculation from the BS and to broadcast the data to NLOS vehicles using LOS relay node. Using RML 1) the mmBS learns about its surroundings through continuous observations 2) the mmBS becomes efficient to predict the relay node earlier when it identifies vehicles at NLOS. The RML achieves 75% of the PDR when the blockage is at its maximum and the results demonstrate the reliability of using Machine Learning in 5G V2X applications. Even though the RML performance is good, it lacks accuracy when it sees rapid changes in traffic pattern. When more number of NLOS nodes is present a single relay may not be suitable to transmit information to all of them especially when the coverage range and the transmission distance is higher. Hence our RML model can be further extended by using Deep RL or Q-learning to accurately predict the temporary blockages and to efficiently select the relay nodes for transmitting data to NLOS vehicles based on the daily changes in traffic pattern especially during the peak hours.

## AUTHORS

**Deepika Mohan** is currently a Master Student at the Department of Electrical and Electronics Engineering, Auckland University of Technology, NZ. Deepika received her previous Master Degree (Communication systems) in India 2011. Her research interest is Machine Learning, mmWave, vehicular communication and 5G.

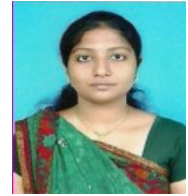

**G. G. Md. Nawaz Ali** (Member IEEE) is working as an Assistant Professor with the Department of Applied Computer Science of University of Charleston, WV, USA. Prior to joining to UCWV, he was working as a post doctoral research fellow with the Department of Automotive Engineering of Clemson University, SC, USA (March 2018 – July 2019). His research interests are in the areas of Vehicular Adhoc networks, Scheduling and Broadcasting and data analytics.

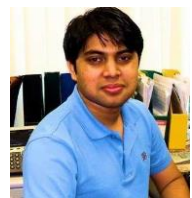

**Peter Han Joo Chong** (Senior member IEEE) is an Associate Head of School (Research) and a Head of Department of Electrical and Electronic Engineering at Auckland University of Technology, Auckland, New Zealand. He received his Ph.D. degrees in Electrical and Computer Engineering from the University of British Columbia, Canada. His research interests are in the areas of wireless/mobile communications systems including radio resource management, multiple access, MANETs/VANETs, green radio networks and 5G-V2X networks. He has published over 200 journal and conference papers, 1 edited book and 9 book chapters in the relevant areas.

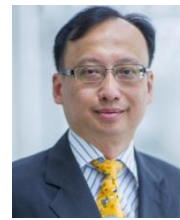